\documentclass{aa}  

\usepackage{graphicx}
\usepackage{txfonts}
\usepackage{natbib}
\bibpunct{(}{)}{;}{a}{}{,}

\begin{document}

   \title{GJ\,1214: Rotation period, starspots, and uncertainty on the optical slope of the transmission spectrum
\thanks{Based on data obtained with the STELLA robotic telescopes in Tenerife, an AIP facility jointly operated by AIP and IAC}}

   \titlerunning{Rotation period of GJ\,1214 and its starspots}

   \author{M. Mallonn\inst{1}, E. Herrero\inst{2,3}, I.G. Juvan\inst{4,5,6}, C. von Essen\inst{7}, A. Rosich\inst{2}, I. Ribas\inst{2}, T. Granzer\inst{1}, X. Alexoudi\inst{1}, K.G. Strassmeier\inst{1}}
   \authorrunning{M. Mallonn et al.}

\institute{Leibniz-Institut f\"{u}r Astrophysik Potsdam, An der Sternwarte 16, D-14482 Potsdam, Germany 
  \email{mmallonn@aip.de}
\and 
Institut de Ci\'{e}ncies de l'Espai (ICE, CSIC), Campus UAB, Carrer de Can Magrans s\/n, 08193 Cerdanyola del Vall\`{e}s, Spain 
\and
Institut d’Estudis Espacials de Catalunya (IEEC), C/Gran Capit\'{a} 2-4, Edif. Nexus, 08034 Barcelona, Spain
\and
Space Research Institute, Austrian Academy of Sciences, Schmiedlstrasse 6, A-8042, Graz, Austria 
\and
Institut f\"ur Geophysik, Astrophysik und Meteorologie, Karl-Franzens-Universit\"at, Universit\"atsplatz 5, A-8010 Graz, Austria 
\and
Institut f\"ur Astro- und Teilchenphysik, Universit\"at Innsbruck, Technikerstrasse 25, A-6020 Innsbruck, Austria 
\and
Stellar Astrophysics Centre, Department of Physics and Astronomy, Aarhus University, Ny Munkegade 120, DK-8000 Aarhus C, Denmark 
}

   \date{Received --; accepted --}

 
  \abstract
   {}
   {Brightness inhomogeneities in the stellar photosphere (dark spots or bright regions) affect the measurements of the planetary transmission spectrum. To investigate the star spots of the M dwarf GJ\,1214, we conducted a multicolor photometric monitoring from 2012 to 2016.}
   {The time-series photometry was analyzed with the light curve inversion tool \texttt{StarSim}. Using the derived stellar surface properties from the light curve inversion, we modeled the impact of the star spots when unocculted by the transiting planet. We compared the photometric variability of GJ\,1214 to published results of mid- to late M dwarfs from the MEarth sample.}
   {The measured variability shows a periodicity of $125\,\pm\,5$~days, which we interpret as the signature of the stellar rotation period. This value overrules previous suggestions of a significantly shorter stellar rotation period. A light curve inversion of the monitoring data yields an estimation of the flux dimming of a permanent spot filling factor not contributing to the photometric variability, a temperature contrast of the spots of $\sim\,370$~K and persistent active longitudes. The derived surface maps over all five seasons were used to estimate the influence of the star spots on the transmission spectrum of the planet from 400~nm to 2000~nm. 
   The monitoring data presented here do not support a recent interpretation of a measured transmission spectrum of GJ\,1214b as to be caused by bright regions in the stellar photosphere. Instead, we list arguments as to why the effect of dark spots likely dominated over bright regions in the period of our monitoring. Furthermore, our photometry proves an increase in variability over at least four years, indicative for a cyclic activity behavior. The age of GJ\,1214 is likely between 6 and 10~Gyr.}
   {The long-term photometry allows for a correction of unocculted spots. For an active star such as GJ\,1214, there remains a degeneracy between occulted spots and the transit parameters used to build the transmission spectrum. This degeneracy can only be broken by high-precision transit photometry resolving the spot crossing signature in the transit light curve. }

   \maketitle
%

\section{Introduction}
For the vast majority of currently known exoplanets, direct spectroscopy is hampered by the strong brightness contrast between the host star and the planet, and by their small angular separation. However, transiting extrasolar planets offer the opportunity to perform indirect spectroscopy of these planets. Planetary spectroscopy using the transit event, the so-called transmission spectroscopy, searches for spectral modifications of the part of the star light shining through the planetary atmosphere during transit. The opacity of the planetary atmosphere is a function of wavelength, dependent on chemical composition \citep{Fortney2010}. Relative differences in opacity can be measured by comparing photometric transit light curves taken at multiple wavelengths \citep{Bean2010,Pont2013,Kreidberg2014,Lendl2016,vonEssen2017,Kirk2017}. At high opacity, the planetary radius appears larger than at wavelengths of weaker opacity \citep{Sing2016,Sedaghati2017,Mallonn2017}.

However, brightness inhomogeneities caused by magnetic activity, starspots or faculae \citep{Strassmeier2009}, modify the derived transit parameters and can mimic spectral features of the planetary atmosphere \citep[e.g.,][]{Oshagh2013,Barstow2015,Herrero2016}. For example, a trend of increasing planetary radius toward blue wavelengths can be caused by scattering properties in the atmosphere of the planet, but also by the presence of starspots on the visible hemisphere of the host star \citep{Pont2013,McCullough2014}. One exoplanet that received a lot of attention in recent years is the super-Earth GJ\,1214b \citep{Charbonneau2009}. Its mean density can be explained by different bulk compositions. The possible scenarios can be distinguished by spectroscopy of the planetary atmosphere \citep{MillerRicci2010}. Because of the relatively large transit depth, such measurements were feasible and conducted by different groups from space \citep[e.g.,][]{Berta2012,Kreidberg2014} and from the ground \citep[e.g.,][]{Bean2010,Croll2011,deMooij2013,Nascimbeni2015}. The planet spectrum was found to be flat from blue optical wavelength to the near-infrared.

The host star GJ\,1214 is an active mid M dwarf of 0.176 solar masses \citep{Anglada2013} that shows photometric variability due to active regions rotating in and out of view \citep{Charbonneau2009}. However, no clear picture of its activity pattern could be derived so far. Several teams performed photometric monitoring campaigns \citep[e.g.,][]{Berta2011,Nascimbeni2015}, but drew different conclusions. Tentative detections of a periodicity in the variability (which is linked to the stellar rotation period) ranges from $\sim\,40$~days \citep{Narita2013} to $\sim\,80$~days \citep{Nascimbeni2015}. \cite{Newton2016b} still lists the stellar rotation period as unknown. The measured amplitude of variability ranges from below 2~mmag in Cousin I \citep{Gillon2014} up to 3\,\% in the Sloan g band \citep{Teske2013}. In principle, this range of results could be partly explained by an insufficiently long monitoring of the host star, and by an unknown evolution of the activity patterns on the host star.

The photometric variability so far was always interpreted as dark spots rotating in and out of view. Starspots on GJ\,1214 have been seen in transit photometry as bumps in several transit light curves \citep{Berta2011,Carter2011,Bean2011,Narita2013,Nascimbeni2015}. A different interpretation of GJ 1214's variability was recently given by \cite{Rackham2017}, who measured an offset toward lower values in the planetary transmission spectrum of the optical data compared to previously published NIR results. They interpreted it as caused by bright regions in the stellar photosphere.

In this work, we want to address the stellar activity and its influence on the transmission spectrum through the analysis of five seasons of multicolor photometric monitoring. GJ\,1214 was observed as part of our monitoring survey VAriability MOnitoring of exoplanet host Stars (VAMOS, Mallonn et al. in prep.) from 2012 to 2016 in the Johnson/Cousins filters B, V, and I. Section \ref{sec_data} describes the observations and the data reduction. In Section \ref{sec_res}, we derive the results for the photometric variability in general and the results obtained by the light curve inversion done with \texttt{StarSim} \citep{Herrero2016}. In Section \ref{sec_disc}, we compare the variability of GJ\,1214 to other mid- to late M dwarfs of the MEarth sample, we discuss the difficulty to extract the optical slope of the planetary transmission spectrum in case of an active host star, and give a discussion on the cyclic activity behavior and the age of GJ\,1214. We finish in Section \ref{sec_conc} with a conclusion.

\section{Observation and data reduction}
\label{sec_data}

The exoplanet host star GJ\,1214 was monitored photometrically over five observing seasons with the robotic telescope STELLA \citep{Strassmeier2004}, located on Tenerife, and its wide-field imager WiFSIP \citep{Granzer2010,Weber2012}. The duration of observations per season and number of acquired images is summarized in Table \ref{obslog}. The data of 2012, and partly of 2013, have already been analyzed and published in \cite{Nascimbeni2015}. They are also included here for an overall analysis. A malfunction of the CCD cooling in the first data points of 2012 was unnoticed in \cite{Nascimbeni2015}. These points are excluded here.

WiFSIP offers a field of view (FoV) of 22\,$'\times$ 22\,$'$ on a scale of 0.32$''$/pixel. The detector is a single 4096$\times$4096 back-illuminated thinned CCD with 15$\mu$m pixels. In 2012, the observations were performed in blocks of five exposures in Johnson V (150~s exposure time) and five exposures in Cousin I (60~s). In 2013, we used Johnson B instead of I because of its higher sensitivity to starspots and observed in blocks of three exposures per filter with 300~s (B) and 150~s (V) exposure time. In 2014 to 2016, the observing blocks consisted of two exposures in B (400~s) and two in V (200~s).

\begin{figure*}
\includegraphics[width=\hsize]{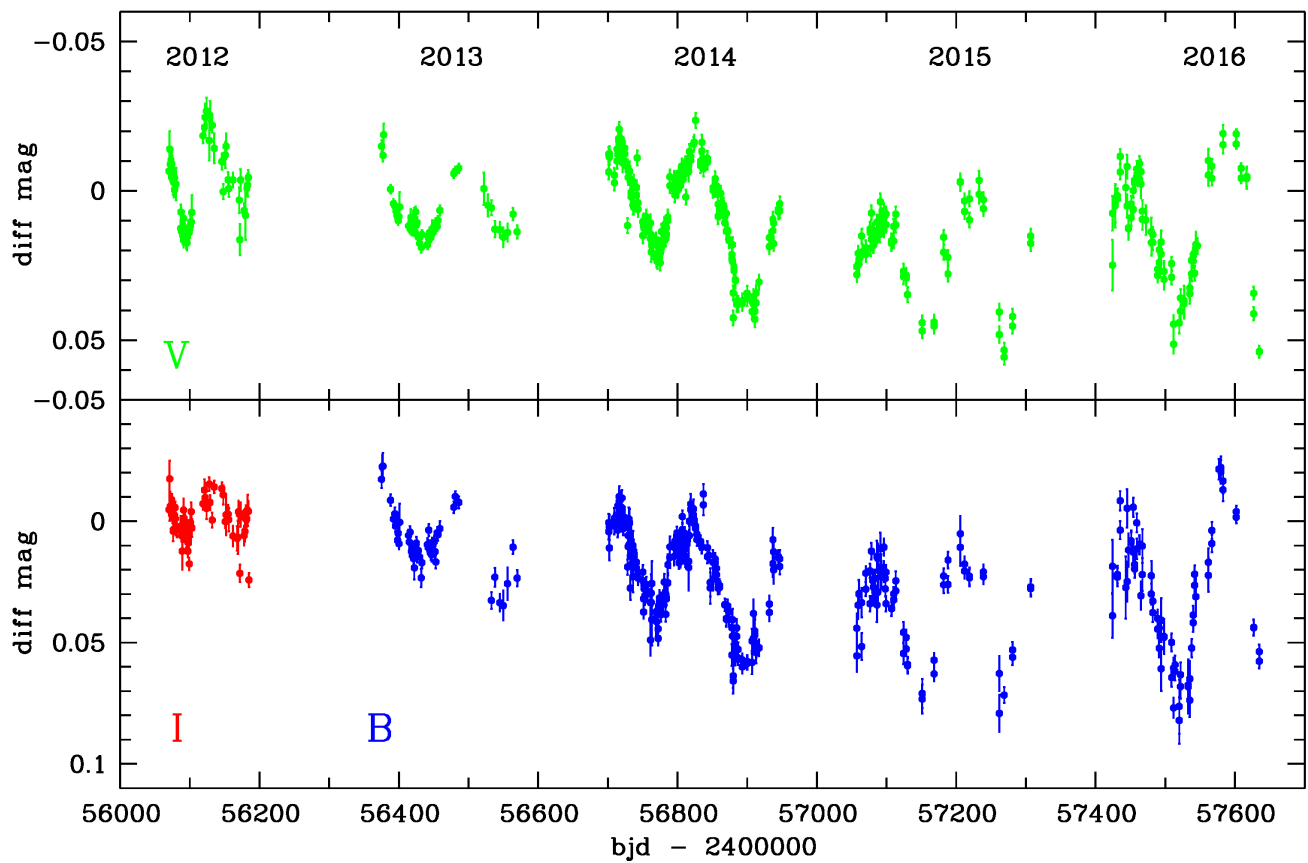}
\caption{Differential light curve of GJ\,1214 covering the time span from 2012 to 2016. Blue data points denote measurements in the Johnson B filter, whereas green and red points indicate observations in the Johnson V and Cousin I filter, respectively.}
\label{plot_lotephot}
\end{figure*}

The data reduction was done with the same ESO-MIDAS routines already used for similar monitoring programs of exoplanet host stars with STELLA WiFSIP \citep[e.g.,][]{MallonnH12,MallonnH19}. Aperture photometry was executed with the publicly available software SExtractor \citep{Bertin96}. 
We employed the SExtractor aperture option MAG\_AUTO, which is an elliptical aperture adjusted individually per frame according to the object light distribution. This option accounts for the changing observing conditions from night to night over the duration of the campaign. For further detail of the data reduction we refer to \cite{MallonnH19} and \cite{Nascimbeni2015}. As comparison stars we used the same four stars within 3$'$ of GJ\,1214 as in \cite{Nascimbeni2015}. We verified that the variability pattern in the derived light curve is not significantly influenced by the choice of the comparison stars. 
We discarded data of lower quality by applying stringent selection criteria which are airmass higher than 1.5, flux less than 50\,\% of average, and unusual values of sky background, width and elongation of point-spread-function, and CCD temperature. From 2013, the criterion of airmass $X\,<\,1.5$ was programmed in the STELLA telescope software that triggers the observations. Therefore, a higher fraction of the observed images met the selection criteria. The data points in 2012 and 2013 were binned per observing block. The resulting light curve is shown in Fig.~\ref{plot_lotephot}.

\begin{table*}
\small
\caption{STELLA monitoring observations of GJ\,1214. The data of 2012 and part of 2013 are presented in \cite{Nascimbeni2015}. The number of data points $n_{\mathrm{data}}$ denotes the value after removal of low-quality data and in 2012 and 2013 after binning the data per observing block.}

\label{obslog}
\begin{center}
\begin{tabular}{cccccc}
\hline
\noalign{\smallskip}
Year              &  2012   &  2013   &  2014   &  2015  & 2016 \\
\noalign{\smallskip}
\hline
\noalign{\smallskip}
Time range        & Mar 21 - Oct 21 &  Mar 24 - Oct 4 &  Feb 13 - Oct 16 &  Feb 4 - Oct 11 & Feb 6 - Sep 3 \\
Indiv. nights & 110  &  81    &  116   &  50  &  53  \\
Filter            & V / I   & B / V  & B / V  & B / V & B / V  \\
n$_{\mathrm{data}}$ & 57  /  72   & 43 / 47      & 206 / 206    & 68 / 76  &  73 / 77 \\
\hline
\end{tabular}
\end{center}
\end{table*}

\section{Results}
\label{sec_res}

\subsection{Stellar variability}
Within the time span of our monitoring, the years 2012 to 2016, there were two other variability measurements of the host star published. \cite{Teske2013} observed multiple transits in 2012 and compared the out-of-transit magnitudes of GJ\,1214 taken in Sloan g band with STELLA WiFSIP in their Fig.~4. The variability pattern in these data agrees well with our measurements from 2012 in the fact that the star was at a similar brightness in late May and late August 2012, while it was dimmer in June. The monitoring of \cite{Narita2013} overlaps the monitoring presented here only by a few days. However, in their Fig.~6, \cite{Narita2013} extrapolated the measured variability pattern to earlier dates. This extrapolation contradicts our measurements as it states a brightness minimum around JD 2456120 for which we measure a maximum in this work.

\begin{figure}
\includegraphics[width=\hsize]{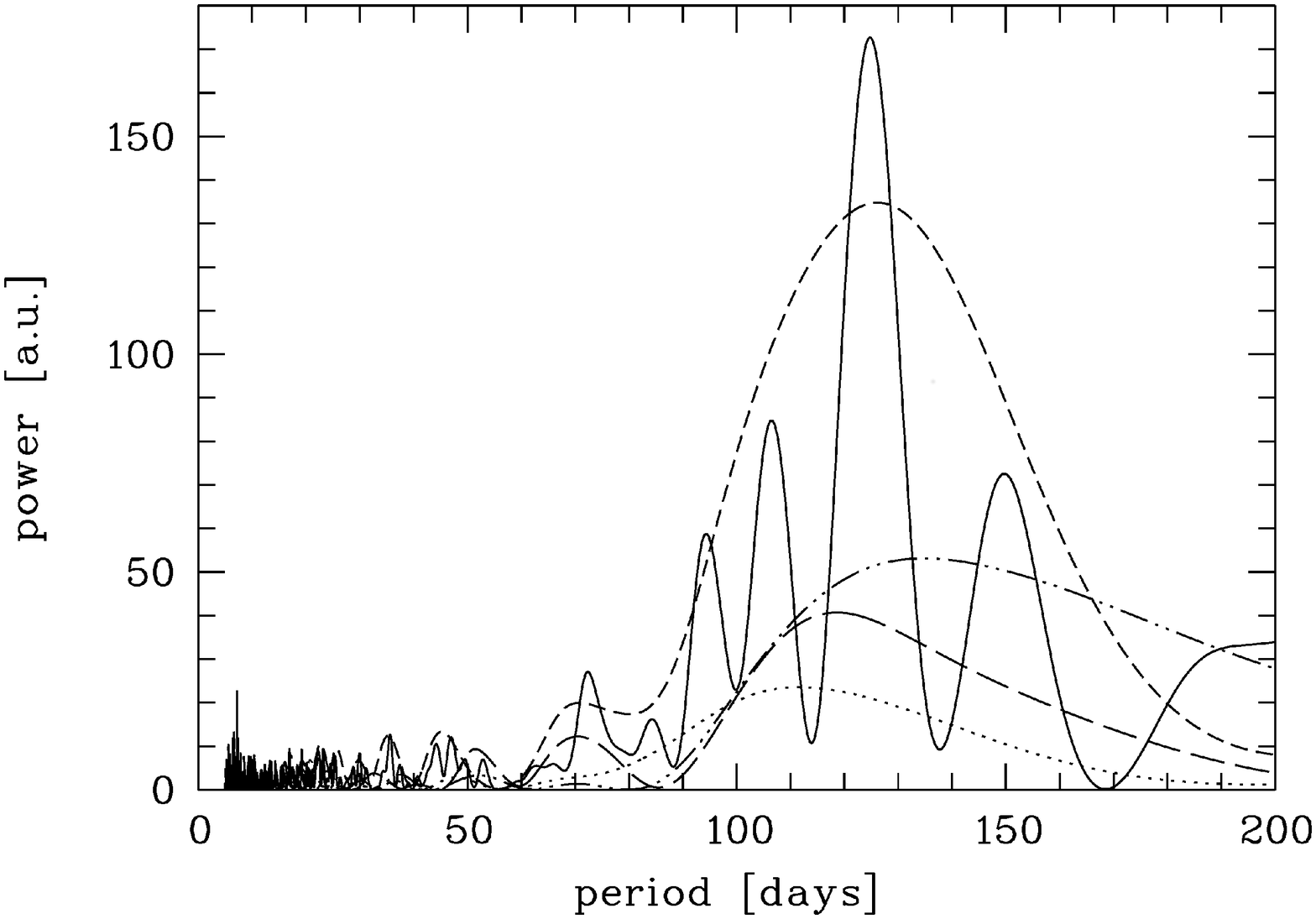}
\caption{Lomb-Scargle periodogram of the differential light curve of GJ\,1214. The black line gives the combined data set B+V from 2014 to 2016, the dotted, short dashed, long dashed, and dashed-dotted lines shows the result for the B+V data of individual years 2013, 2014, 2015, and 2016, respectively.}\label{plot_periodo}
\end{figure}

We computed a Lomb-scargle periodogram for each season individually and for combined seasons. Here, we merged the data sets of the available colors. The four seasons 2013 to 2016 B+V consistently showed a signal P$\,>100$~days, whereas the periodogram of season 2012 reaches peak power at P$\,\sim\,70$~days as already shown in \cite{Nascimbeni2015}. We obtain the most significant signal in the periodogram for the data B+V 2014 to 2016 at a period of $125\pm5$~days (Figure \ref{plot_periodo}). The uncertainty is the HWHM of the peak in the periodogram.

Our STELLA B and V band data measure a full amplitude variability of 5.3\,\% in V and B, which is the highest variability measured so far for this exoplanet host star. Other papers containing long-term photometry of GJ\,1214 from 2010 to 2012 already showed a rise in variability amplitude. In 2010, \cite{Berta2011} measured a full-amplitude in the V band of 1.3\,\%. \cite{Nascimbeni2015} reported an increase in amplitude from their analysis of the 2012 STELLA data to about 2\,\% in I. 
For the same year 2012, \cite{Teske2013} obtained an out-of-transit variability of about 3\,\% in the bluer Sloan g band. We discuss the increase in amplitude of photometric variability from the year 2008 to 2016 as indicators for a stellar activity cycle in Section \ref{chap_cycle}.

\subsection{Spot modeling with \texttt{StarSim}}
\label{spotmodel}

The host star GJ\,1214 is known to present stellar variability due to inhomogeneities in the photosphere \citep{Berta2011,Nascimbeni2015}. Time series multicolor photometry allows for the determination of the amount and longitude distribution of active regions on the stellar surface. From this, it is also possible to derive the temperature contrast of apparent starspots in the photosphere \citep{Strassmeier2009}.

Among the different spot modeling techniques, we chose the \texttt{StarSim} tool \citep{Herrero2016} to analyze the data described in Section \ref{sec_data}. The methodology consists of the surface integration of fluxes using Phoenix BT-Settl atmosphere models \citep{Allard2011}. Different temperatures were considered for the spots and for the quiet photosphere. The total flux produced by the visible stellar surface was obtained by adding the contribution of the visible surface elements, considering the limb darkening computed from Kurucz ATLAS9 models \citep{Castelli2004}. \texttt{StarSim} models the surface of an active star by considering a certain amount of cool spot elements. The number of active surface elements, as well as their locations and lifetimes, are free parameters of the model. This number corresponds to the amount of independent regions, each one with a particular position and size, that are used by \texttt{StarSim} to model-fit the data. The \texttt{StarSim} fitting routine may group several of those elements on a single active region or spot group, so the fixed number of active surface elements does not determine the final number of active regions, but the maximum. While increasing the number of active elements may in principle improve the solution, constraining this parameter is a way of regularizing the solution and avoid fitting the noise present in the data. A certain amount of bright faculae around cool spot elements can also be included in the model, considering a hotter surface temperature and a limb brightening law as described by \cite{Meunier2010}. The best solution for the stellar surface map is found by means of a simulated annealing approach \citep{Herrero2016}. The parameter space for some of the physical properties of the system can also be explored using \texttt{StarSim} and MCMC techniques (Rosich et al. in prep.). These were not used in the this work because of the computational cost of performing a large number of simulations of long time series, especially when the surface map is complex, as is the case for GJ1214. Instead, a more simple grid searh method was used. Finally, in the current version of \texttt{StarSim}, multiple time series data can be combined to provide a single solution for the stellar parameters and the surface map.

\begin{table*}
\small
\caption{System parameters of GJ\,1214.}
\label{tab:params}
\begin{center}
\begin{tabular}{lcccc}
\hline
\noalign{\smallskip}
Parameter & Symbol & Value & Unit & Reference \\
\noalign{\smallskip}
\hline
\noalign{\smallskip}
Effective temperature &  $T_{\rm eff}$ & 3252$\pm$20 & K & \cite{Anglada2013} \\
Temperature contrast of spots & $\Delta T_{\rm spots}$ & $372_{-182}^{+138}$ & K & this work, fitted \\
Stellar surface gravity & $\log g$ & 5.01$\pm$0.07 & - & \cite{Anglada2013} \\
Facula to spot area ratio & $Q$ & 0.0 & -  & this work, fixed \\
Rotation period & $P_{\rm rot}$ & 125$\pm$5 & days & this work, fitted \\Planet orbit inclination & $i$ & 88.80$_{-0.20}^{+0.25}$ & $^{\circ}$ & \cite{Berta2011} \\
Planet orbital period & $P_{\rm planet}$ &  1.58040482$\pm$0.00000024 & days & \cite{Kreidberg2014} \\
Planet transit epoch & $t_{\rm 0}$ & 54966.52488$\pm$0.00004 & days & \cite{Kreidberg2014} \\
Planet impact parameter & $b$ & 0.19$^{+0.08}_{-0.11}$ & - & \cite{Berta2012} \\
Planet to star radius ratio & $k$ & 0.1160$\pm$0.0005 & - & \cite{Berta2012} \\
Johnson B photometric zero point & $z_{\rm B}$ & $1.06_{-0.04}^{+0.05}$ & - & this work, fitted \\
Johnson V photometric zero point & $z_{\rm V}$ & $1.06_{-0.04}^{+0.05}$ & - & this work, fitted \\
Johnson I photometric zero point & $z_{\rm I}$ & $1.04_{-0.04}^{+0.05}$ & - & this work, fitted \\
\hline
\end{tabular}
\end{center}
\end{table*}

In this work, we fit for the surface map of GJ\,1214 for the five seasons covered by the data described in Section \ref{sec_data} using \texttt{StarSim} and the parameters shown in Table \ref{tab:params}. In \texttt{StarSim}, the limb darkening for each surface element is computed by interpolating intensities from Kurucz ATLAS 9 models. As these are truncated at 3500~K, the intensities for GJ~1214 need to be extrapolated. However, the quadratic coefficients computed with \texttt{StarSim} for the three analyzed filters are in good agreement with the coefficients computed by \cite{Claret2011} using Phoenix models.

The stellar axis inclination is an input parameter for \texttt{StarSim}'s simulations. As GJ\,1214 is a very slow rotator, there are no well resolved $v\sin i$ measurements or observations of the Rossiter-McLaughlin effect. In this work we assume that the planet - star system is aligned, as is the case in other super-Earths \citep{Barnes2015}. Given that we fix the stellar axis inclination of GJ\,1214 to 88.80$_{-0.20}^{+0.25}$ \citep{Berta2011}, that is, nearly equator on, the solutions obtained with \texttt{StarSim} for the stellar surface map are degenerated in latitude, and only the recovered longitudes of the spot distribution and the projected filling factor of spots can be considered reliable.

\begin{figure}
\includegraphics[width=\hsize,angle=0]{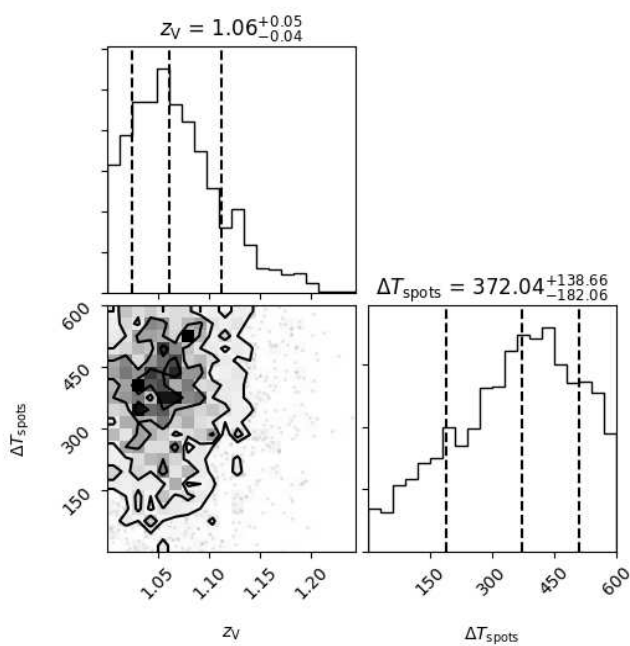}
\caption{Probability distribution of the photometric zero point in the Johnson V band ($z_{\rm V}$) and the temperature contrast of the spots ($\Delta T_{\rm spots}$). The vertical dashed lines indicate the median and the 68\,\% confidence limits of the distributions.}
\label{GJ1214_corner}
\end{figure}

In a first step of the modeling process, we performed a grid search by running 10000 solutions considering a simple surface model consisting of four active surface elements. This is done in order to find the best fit for the photometric zero points ($z_{\rm B}$, $z_{\rm V}$ and $z_{\rm I}$) and the temperature contrast of the spots ($\Delta T_{\rm spots}$). The choice of four active elements in the first step of the \texttt{StarSim} modeling was made after trying solutions with two to ten elements, and finding that four was the minimum which successfully reproduced the light curve by visual inspection. While this is a first suggestion of the number of long-lived active regions in GJ~1214, the detailed evolution of the stellar surface map will be discussed after the second step of the modeling procedure.
The photometric zero points are defined as $z_{i}=1/\langle f_{i}\rangle$, where $f_{i}$ is the relative flux for filter $i$. A relative flux $f_{i}=1$ for $i=B,V,I$ is assumed for a completely quiet photosphere. A uniform grid of 100x100 different values for $\Delta T_{\rm spots}$ and $z_{\rm V}$ is defined within the ranges [0 K, 600 K] and [1.0, 1.25], respectively. Then, the $\chi^2$ for the best surface map solution is computed in each case considering all the multicolor information. The rest of the photometric zero points ($z_{\rm B}$ and $z_{\rm I}$) are dependent on the two free parameters, $\Delta T_{\rm spots}$ and $z_{\rm V}$. Therefore, these are used to compute $z_{\rm B}$ and $z_{\rm I}$ for each solution by simulating light curves with \texttt{StarSim} in B and I filters, considering the whole time span of the observed datasets. A weight scaled with $\sim\exp[-(\chi^2/\chi^2_{\rm scale})]$, with $\chi^2_{\rm scale}=100$, is applied to each solution in order to produce the distributions displayed in Figure~\ref{GJ1214_corner}. The purpose of such scaling is to define a distribution which presents a peak around the solutions with the lowest $\chi^2$. The median and the 68\,\% confidence limits of the distributions (see Figure~\ref{GJ1214_corner}) are adopted as the best results for $\Delta T_{\rm spots}$ and $z_{\rm V}$. 
The Phoenix model grids used in \texttt{StarSim} are truncated at 2600~K, thus preventing us to explore solutions for $\Delta T_{\rm spots}>600$~K in the case of GJ\,1214. However, the result for the temperature contrast of the spots is in good agreement with the range of values observed for other M dwarfs \citep{Berdyugina2005}. Instead of the more commonly used MCMC techniques, this type of analysis was chosen because of the computational cost of performing thousands of simulations with \texttt{StarSim} \citep[see][]{Herrero2016}, given the length of the photometric time series and the complexity of the surface of GJ\,1214. The results are shown in Table \ref{tab:params}. 

When multicolor time-series photometry is available, the variability amplitude differences allow for a measurement of the average temperature contrast of the active regions. The results from the first step of the \texttt{StarSim} modeling already show that the photometry can be modeled considering only cool spots with a temperature contrast around $\sim 400$~K. Therefore, while a certain number of faculae might be present in the active regions, our results show that the photometric variability on GJ1214 is dominated by the effects of cool spots. This is also supported by the fact that, while transit photometry of GJ\,1214b has shown multiple spot-crossing events \citep{Berta2011,Carter2011,Nascimbeni2015,Narita2013,Kreidberg2014}, no signs of faculae features have been reported yet. Also, \cite{Youngblood2017} detected emission from GJ1214 at certain metal lines and also in Ca~K, but among the 15 M dwarfs of the \texttt{MUSCLES} survey \citep{France2016,Youngblood2017}, GJ1214 is at the lower end of UV emission line flux and it shows one order of magnitude less emission line flux in Ca~K than the rest of the targets. In order to avoid additional degeneracies in the models, in this work we assumed that the presence of faculae or bright regions is negligible, so the facula-to-spot area ratio is fixed to $Q=0$.
In a recent paper \cite{Rackham2017} suggest that GJ\,1214b transmission spectroscopy observations are modified by unocculted bright regions in the stellar photosphere. We discuss this further in Sect.~\ref{disc_transm}.

\begin{figure*}
\includegraphics[width=\hsize,angle=0]{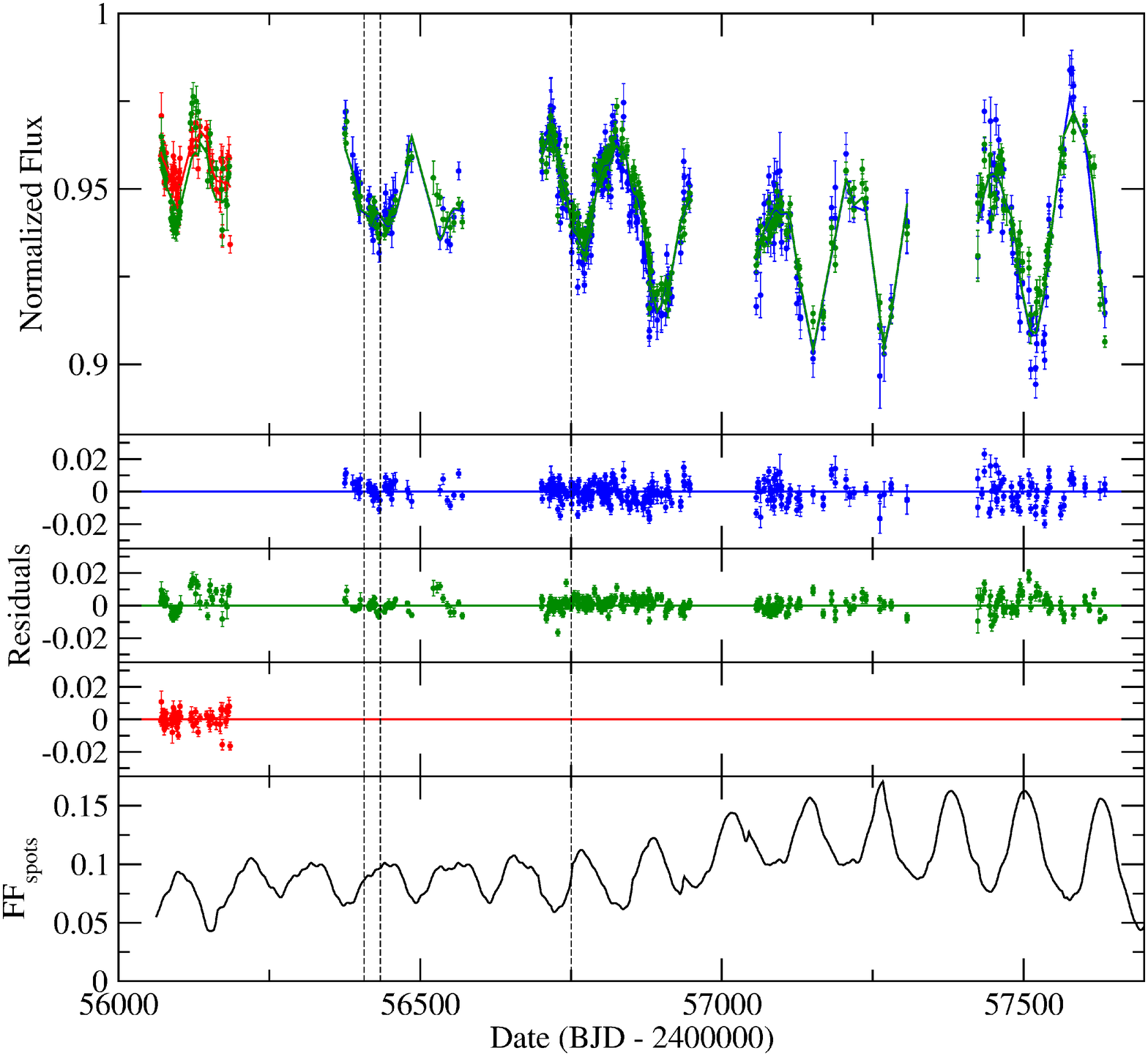}
\caption{Top panel: Light curves of GJ\,1214 from 2012 to 2016. Blue data points denote measurements in the Johnson B filter, whereas green and red points indicate observations in the Johnson V and Cousin I filter, respectively. The photometric zero points are set according to the results presented in Table~\ref{tab:params}. The light curve models obtained with \texttt{StarSim} for the three bands are plotted with solid lines. Middle panels: Residuals between the data and the model for the B, V, and I filter (from top to bottom). Bottom panel: Evolution of the projected filling factor of spots for the best fitting model obtained with \texttt{StarSim}. The three vertical dashed lines in all panels indicate the mid times of the transits analyzed by \cite{Rackham2017}.}
\label{GJ1214_starsim}
\end{figure*}

In a second step of the fitting procedure with \texttt{StarSim}, we fixed the parameters from Table~\ref{tab:params}, including the starspot temperature contrast and the zeropoints found in the first step. So we only fitted for the number, the lifetimes, and the surface distribution of the active elements. Figure \ref{GJ1214_starsim} displays the relative fluxes for the STELLA datasets in B, V and I filters, together with the best fit found with StarSim. The standard deviations of the residuals, in relative flux units, are $\sigma_{\rm B}=3.71 \cdot 10^{-3}$, $\sigma_{\rm V}=2.50 \cdot 10^{-3}$ and $\sigma_{\rm I}=3.36 \cdot 10^{-3}$ in B, V and I bands respectively. These are near 1.7 times the standard error of the mean points for the three light curves. Figure~\ref{GJ1214_starsim} also displays the evolution of the projected filling factor of spots given by our best solution.

\begin{figure}
\includegraphics[width=\hsize,angle=0]{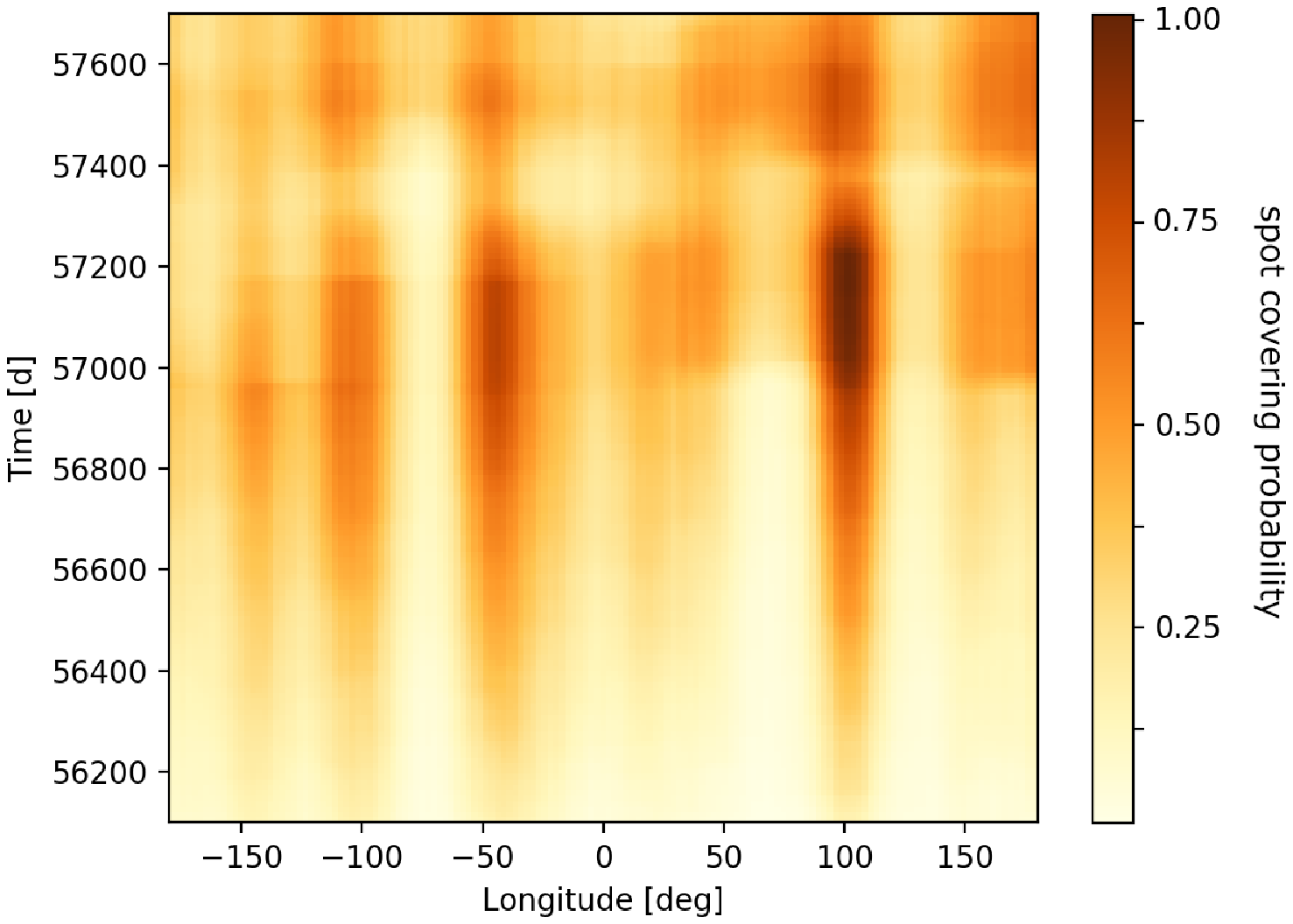}
\caption{Spot covering probability, displayed with different colour intensities from white (quiet photosphere) to dark brown (cool spots), versus time and longitude, for the average of 256 \texttt{StarSim} surface map solutions with the parameters fixed from Table~\ref{tab:params}.}
\label{GJ1214_map}
\end{figure}

In order to estimate the properties of active regions during the time span covered by our data, the parameters from Table~\ref{tab:params} were adopted and 256 solutions for the surface map were run with \texttt{StarSim}. The average map, projected to the longitude axis, is shown in Fig.~\ref{GJ1214_map}, displaying the spot coverage probability vs. longitude and time. The light curves and also the projected filling factor of spots produced by these solutions will be nearly identical, while the surface maps can be slightly different given the existing degeneracies with the size and latitude of the spots. By averaging 256 solutions we obtained the surface map of the spot covering probability, preventing small short-lived spots from appearing on it. This is also a way of regularizing the solution and obtaining a smoother surface map that puts the focus on the main regions where spots are preferred by the model. The map displayed in Fig.~\ref{GJ1214_map} 
shows that there are at least four long lived active regions, including a dominant one which persists during all the time span covered with our photometry (more than four years). Also, all the active regions are stable in longitude in the reference frame of the stellar rotation period, indicating no signs of differential rotation.

\subsection{Spectral signature of active regions}
\label{sect:transm}

The atmospheres of exoplanets are often studied by using transmission spectroscopy techniques, in which low resolution spectroscopy is obtained during planetary transits in order to simultaneously measure the transit depth at multiple wavelengths. A slope toward the blue optical wavelengths seems to be present for most gas giants \citep{Sing2016, Mallonn2017}. A widely accepted interpretation is the presence of a cloud or haze layer \citep{Wakeford2015,Pinhas2017}. However, a unique physical interpretation of the spectral slope does not exist because spots on the host star can mimic the slope as well as third light from another stellar source along the line of sight \citep{McCullough2014,Herrero2016}.

In this work, we used the results from the modeling of the photometric time series described in Sect.~\ref{spotmodel} in order to analyze a simulated series of transits of GJ\,1214b for the potential effect of the starspots on the planetary transmission spectrum. The procedure is analog to the one shown for HD189733~b by \cite{Herrero2016}. While the degeneracies in our surface map solution, especially affecting the latitudes of the spots, prevent us from simulating specific transit events including all the particular effects of occulted and unocculted spots, we were able instead to simulate a large series of transits and analyze the statistics of the transmission spectra affected by a wide diversity of spot distributions. In the current work, the best solution for the stellar surface map, together with the stellar parameters listed in Table~\ref{tab:params}, were considered in order to generate series of simulated transits including the effects of cool spots on the stellar surface. Each series contains the 635 transits occurring during the time intervals covered by the photometry in Fig.~\ref{GJ1214_starsim}. An additional transit was simulated considering the star with a completely unspotted surface with the purpose of computing precise limb darkening coefficients for each wavelength bin. A series of light curves was simulated for each transit event, covering the wavelength range from 400~nm to 2000~nm in bins of 50~nm, so a transmission spectrum with 32 points is produced for each transit. The sections of the transits affected by spot crossing events were removed from the simulated light curves. This is done automatically by \texttt{StarSim} as an option when generating transit light curves, by using the information about the relative position between the planet and the active regions at any time of the simulation.

\begin{figure}

\includegraphics[width=\hsize,angle=0]{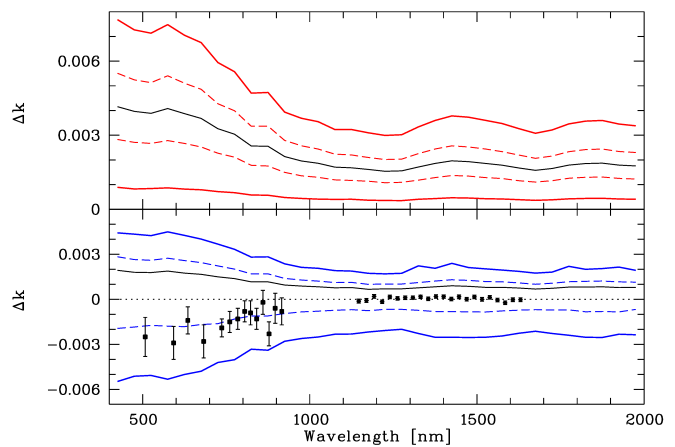}
\caption{Modifications of the planet to star radius ratio by spots. The upper plot shows the variations ($\Delta k=k-k_{\rm unspotted}$) with wavelength, which we derived from analyzing the 635 simulated transits in 32 wavelength bins. The black line indicates the mean values of $\Delta k$ for each wavelength bin with the dashed red line indicating the 68\,\% confidence interval. The solid red lines indicate the region between the minimum and maximum $\Delta k$. In the bottom plot, we show the same for the 278 transits occurring during seasons 2013 and 2014, without removing spot crossing features. The blue lines indicate the region between the minimum and maximum $\Delta k$. The overplotted optical measurements are from \cite{Rackham2017}, the near-IR data are from \cite{Kreidberg2014}.}
\label{GJ1214_tdv}
\end{figure}

The simulated series of transits were analyzed using \texttt{JKTEBOP} \citep{Southworth2004,Southworth2008}. A square root limb darkening law was adopted, as this is known to be the optimal approximation for late type stars in the near infrared \citep{vanHamme1993}. Limb darkening coefficients, together with the sum of the radii and the planet to star radius ratio ($k=R_{\rm p}/R_{\rm s}$), were kept as free parameters when fitting the unspotted transit of each passband. For all the simulated passbands, the computed limb darkening coefficients are in good agreement with those obtained by \cite{Claret2000} for commonly used optical and near infrared filters. Then, we fixed these coefficients for fitting the rest of the transit series, thus deriving the variations of $k$ produced by the presence of unocculted spots. For each transit $i$, the transit depth variation $\Delta k_{\rm i}=k_{\rm i}-k_{\rm unspotted}$ was obtained, with $k_{\rm unspotted}$ being the planet to star radius ratio of the unspotted transit, adopted as reference. Finally, we computed the mean and standard deviation of $\Delta k$ for the 635 transits in each wavelength bin. These are displayed in Fig.~\ref{GJ1214_tdv}, thus showing the chromatic signature of spots on transit depth measurements. Such signature is strongly dependent on the amount of spots and their temperature contrast ($\Delta T_{\rm spots}$). Therefore, only by simultaneously analyzing chromatic information (i.e., multiple wavelength photometry) we could recover accurate properties of spots and their chromatic effects on transmission spectroscopy measurements. Mean $\Delta k$ values and 68\% confidence levels are also plotted in Fig.~\ref{GJ1214_tdv} for all the analyzed spectral range. The spectral signature of spots on $\Delta k$ was also computed for the transits occurring in seasons 2013 and 2014 without removing the effects off spot crossing events from the simulated light curves. This is displayed in the bottom panel of Fig.~\ref{GJ1214_tdv}. The bumps introduced by occulted spots in the transit light curves make the distribution of $\Delta k$ to deviate from a Gaussian, thus causing the mean $\Delta k$ values to be close to the upper 68\% confidence level.

According to our results from simulating transit depth variations, the spots present on the surface of GJ\,1214 can bias the planet to star radius ratio by up to $7\cdot 10^{-3}$ in the blue optical wavelengths and up to $3\cdot 10^{-3}$ in the near infrared. In all our simulations where spot occultation effects were removed there is a positive $\Delta k$ due to a permanent spot coverage in GJ\,1214 (see Fig.~\ref{GJ1214_map}). In transmission spectroscopy, we are interested in the relative change of $\Delta k$ over wavelength instead of its absolute value. This relative change would manifest as a spectral slope. At the maximum of the spot filling factor in the year 2015 it amounts to an amplitude of $(4.3\pm1.6)\cdot 10^{-3}$ between 500 and 1200~nm, while for an average filling factor between 2012 and 2016 it amounts to $(2.5\pm0.9)\cdot 10^{-3}$. This is a factor of 2.5 above the typical uncertainty of a measurement of $k$ at blue wavelengths \citep{deMooij2013,Nascimbeni2015}. Therefore, the influence of unocculted spots in the optical is not negligible during a phase of large photometric variations. We note that the majority of transit observations for transmission spectroscopy of GJ\,1214b have been taken from 2009 to 2012, where the star might have had a significantly lower spot coverage due to a magnetic activity cycle (see Section \ref{chap_cycle}). The most precise transmission spectrum of GJ\,1214b was derived from 12 transit observations with the HST by \cite{Kreidberg2014}, observed between September 2012 and August 2013. Our simulations show that their covered wavelength range from 1100 to 1700~nm is generally less affected by spots and shows only smooth variations.

\section{Discussion}
\label{sec_disc}

\subsection{Variability in the context of MEarth M dwarfs}
\begin{figure}
\includegraphics[width=\hsize,angle=0]{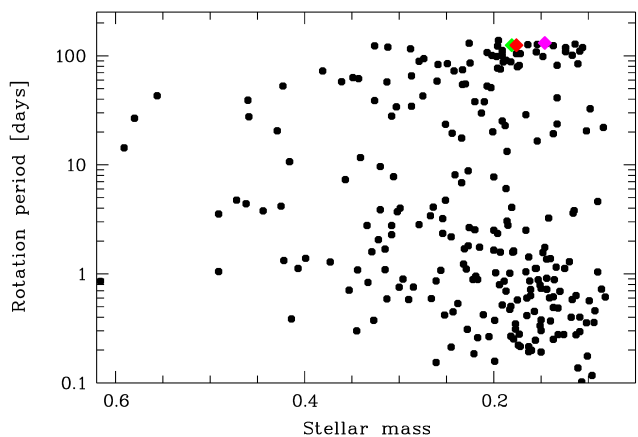}
\caption{Mass-period diagram of MEarth targets. Black data points are taken from \cite{Newton2016a}, GJ\,1214 is shown in red, GJ1132 and LHS1140 in green and purple, respectively. The rotation period of GJ\,1214 of $125\,\pm\,5$~days matches the upper envelope of the distribution of increasing rotation periods with increasing stellar mass.}
\label{plotmassper}
\end{figure}

The exoplanet of GJ\,1214 was discovered by the ground-based transit survey MEarth, which monitors several thousands of mid-to-late M dwarfs to search for transiting planets \citep{Berta2012b}. \cite{Newton2016a} analyzed the variability amplitude and periodicity of 387 M dwarfs of the MEarth northern target list. To place GJ\,1214 in this context, we compared the rotation period and variability amplitude derived here for GJ\,1214 with those of the Newton sample, which did not include our target. In Figure \ref{plotmassper} we reproduce the mass-period diagram, showing that lower-mass stars spin down to longer rotation periods than higher-mass stars, a result already reported in \cite{Irwin2011} and \cite{McQuillan2014}. The rotation period of GJ\,1214 of 125 days is among the longest periods found, but is well within the upper envelope of detected rotation periods. The longest periods found by \cite{Newton2016a} are about 140 days. GJ\,1214 covers the same parameter space in this diagram as the other two planet host stars found by the MEarth survey, GJ\,1132 \citep{Berta2015} and LHS\,1140 \citep{Dittmann2017}.  

\begin{figure}
\includegraphics[width=\hsize,angle=0]{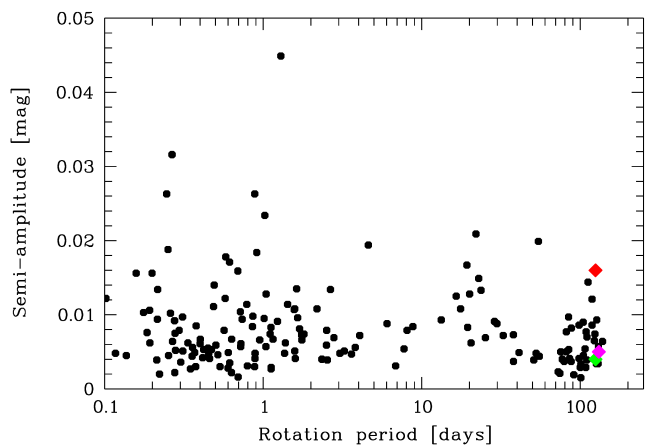}
\caption{Rotation-amplitude diagram of MEarth targets of masses below 0.25~$\mathrm{M}_{\odot}$. Black data points are taken from \cite{Newton2016a}, GJ\,1214 is shown in red, GJ\,1132 and LHS\,1140 in green and purple, respectively. GJ\,1214 is one of the most variable stars among the group of very slow rotators. The result of \cite{Newton2016a} of no correlation between rotation and photometric variability amplitude is strengthened. }
\label{plotrotamp}
\end{figure}

In the next step, we reproduced the period-amplitude diagram of stars less massive than 0.25 solar masses of \cite{Newton2016a}. For GJ 1214 we transformed the peak-to-peak variability amplitude of 5.3\,\% in Johnson V to a semi-amplitude in the MEarth filter of about 1.6\,\%. This transformation was done by simulating the light curve from 2012 to 2016 in the MEarth passband with \texttt{StarSim}, using the parameters found in our analysis (see Table~\ref{tab:params}). The overall system response of an MEarth telescope is taken from \citep{Nutzman2008}, the MEArth filter resembles approximately a combination of Sloan i and z. In the period-amplitude diagram Figure \ref{plotrotamp} we see GJ\,1214 to be among the most variable long-period M dwarfs, strengthening the result of \cite{Newton2016a} that there is no anti-correlation of period and variability amplitude for low-mass M dwarfs. GJ\,1132 and LHS\,1140 show lower photometric variability by a factor of three. We note that \cite{KadoFong2016} lists several low-mass M dwarfs with rotation periods longer than 100 days and even higher variability amplitudes than GJ\,1214 in the slightly redder Pan-STARRS1 z-band.

In HARPS spectra used for radial velocity measurements in the discovery paper \citep{Charbonneau2009}, no activity induced chromospheric emission in either the H$\alpha$ or the NaI doublet was detected. Very weak emission in Ca II K was detected by \citep{Youngblood2017}. \cite{Newton2017} derived a relation between stellar rotation period and chromospheric activity for M dwarfs of the MEarth sample, by which chromospherically inactive stars of 0.15 solar masses should have rotation periods of longer than 80 days. This derivation contradicts previously estimated rotation periods of GJ\,1214 of 40 or 53 days \citep{Narita2013,Berta2011}. Our new derivation of a stellar rotation period of $125\,\pm\,5$~days agrees well with this activity-rotation relationship. In the same work, \cite{Newton2017} described a correlation between chromospheric activity and amplitude of photometric variability. The spectroscopic HARPS observations from \cite{Charbonneau2009} were taken in the year 2009, when GJ\,1214 was at a state of lower photometric variability (see Section \ref{chap_cycle}). Spectroscopic observations of the year 2014 or later would be needed to verify if this activity-variability trend also applies to GJ\,1214. According to this, the increase in photometric amplitude measured in this work might be accompanied by an increase in chromospheric activity.

\subsection{An activity cycle for GJ\,1214?}
\label{chap_cycle}

The magnetic activity of the Sun undergoes a cyclic behavior which observationally manifests in the $\sim\,11$~yr sunspot cycle. Observations of F to early M dwarfs indicate that a significant fraction of these solar-type stars displays similar cycles \citep{Olah2009,Strassmeier2009,SuarezMascareno2016}. Stars later than spectral type $\sim$\,M3.5 are expected to have fully convective interiors and therefore a different dynamo process responsible for magnetic activity than the earlier type solar-like stars \citep{Charbonneau2014}. The question of whether or not fully convective stars show a cyclic behavior of magnetic activity is still open, but recent observational work points toward this possibility. The M6 dwarf Proxima Centauri shows an approximately seven year photometric variability indicative of an activity cycle \citep{Wargelin2017}. In another work, \cite{Hosey2015} present four fully convective M dwarfs with cyclic photometric behavior from four to eight years period.  \cite{SuarezMascareno2016} found cycles from four to ten years for nine M4 to M6 dwarf stars. 

The lower panel of Figure \ref{GJ1214_starsim} shows an increase of the spot filling factor for GJ\,1214 from 2012 to 2015. Additionally, we see a steady increase in amplitude of variability in our STELLA data, starting from a full-amplitude in V of $\sim$\,2.5\,\% in 2012 up to 5.3\,\% in 2016. Transformed to the MEarth filter, these two values correspond to 1.4\,\% and 3.2\,\% of full amplitude. \cite{Berta2011} presented MEarth photometry from 2008 to 2010 with photometric full amplitudes of below 1\,\%. 
For the years prior to our monitoring campaign, it is unclear whether the increase in variability amplitude of GJ\,1214 is driven by an overall increase in activity, potentially caused by a cyclic behavior of the stellar activity. Another reason for a larger variability amplitude might be a change in spot distribution toward less homogeneity compared to previous years. The shorter periods derived by previous studies might indicate the presence of more than one dominating spot group in 2010 and therefore a more homogeneous spot distribution than in 2013 to 2016. The \texttt{StarSim} light curve modeling of the STELLA data in this work suggests no significant changes in the spot distribution from 2012 to 2016. Instead, several active longitudes remained stable over most of the observed time, but the spots grew in size. Therefore, the increase in variability in the STELLA data can be interpreted as a cycle of more than four years length. Solely comparing the photometric variability amplitude from 2008 to 2016, it appears that the cycle might be longer than eight years.

\subsection{On the age of GJ\,1214}
The M dwarf GJ\,1214 shows no sign of youth, for example, it shows no or very weak chromospheric line emission \citep{Charbonneau2009,Youngblood2017}. Low-mass stars spin down as they age, thus in principle we can use the stellar rotation period derived in this work as an age indicator. However, the quantitative framework exploiting this effect called gyrochronology \citep{Barnes2003,Barnes2010} has not yet been empirically anchored for fully convective M dwarfs. Therefore we only use qualitative arguments here. \cite{Newton2016a} use age-space velocity relationships to estimate a mean age for the population of slowly rotating solar neighborhood mid M dwarfs of 5$_{-2}^{+4}$~Gyr. The Alpha Cen system is believed to be somewhat older than the Sun, about 5.6~Gyr old \citep{DeWarf2010}, harboring the mid M dwarf Proxima Centauri with a rotation period of 83~days \citep{Benedict1998}. Barnard's star is suggested to be a very old, thick disk star, potentially exceeding 10~Gyr \citep{Choi2013} with a tentatively detected long stellar rotation period of about 130 days \citep{Benedict1998}. GJ\,1214 is likely to be older than Proxima Centauri and as old or slightly younger as Barnard's star. Thus we conclude GJ\,1214 to have an age of between 6 and 10~Gyr.

\subsection{Optical slope in the planetary transmission spectrum}
\label{disc_transm}
\cite{Rackham2017} observed three transit events in the years 2013 and 2014 and found an offset in the spectrum between optical and near-infrared wavelengths, in contrast to the optical results of \cite{Nascimbeni2015}, \cite{deMooij2013}, and \cite{Narita2013}. The observations for the latter three works were taken in the years 2011 and 2012. The tentative downward slope toward shorter wavelengths is consistently present in all three data sets of \cite{Rackham2017}, and their favored interpretation are modifications of a flat planetary spectrum by unocculted bright regions in the stellar photosphere. In this interpretation, the stellar photosphere is either covered only with bright regions without dark spots, or the effect of the bright regions dominates. Here, we provide arguments for why we consider this scenario to be unlikely.

As presented in Sect.~\ref{spotmodel}, the light curve inversion resulted in active regions that are around 400~K cooler than the quite photosphere. The long-term photometry of this work can be modeled solely by dark regions without any need for a more complex mixture of dark and bright regions. The temperature contrast between quiet photosphere and active surface elements is mainly derived by the multicolor information of the long-term variability, and the light curve inversion provides no indications for bright regions dominating this variability.

There is evidence of dark spots on the stellar photosphere of GJ\,1214 by spot crossing events in transit observations from 2009 to 2013 \citep{Berta2011,Carter2011,Nascimbeni2015,Narita2013,Kreidberg2014}, and 2017 (Juvan et al. in preparation). The spot crossing events in two transits of \cite{Kreidberg2014} from August 2013 happened chronologically in between the transit observations of \cite{Rackham2017} in 2013 and 2014. The dominating spot groups are long-living (see Sect.~\ref{spotmodel}), thus it is reasonable to assume that spots were always present in the photosphere of GJ\,1214 between 2009 and 2017. However, none of the numerous published transit light curves showed evidence for a crossing of bright regions.

In a very recent study, Juvan et al. (in preparation) analyzes 25 transit light curves observed in the year 2017 together with simultaneous photometric monitoring. Spot crossing events have been found in the transit light curves only at times of a relative brightness dimming of the host star, indicative for more spots on the surface at times when the star is faint and therefore a spot dominated stellar variability.

The transit observations of \cite{Rackham2017} overlap with the photometric monitoring presented here. The epochs of the three transits are marked as vertical dashed lines in Figure~\ref{GJ1214_starsim}. At each transit, the star is near a minimum of relative flux. Hence, to be potentially responsible for the measured transmission spectrum, bright regions had to dominate over dark regions even at times of stellar flux minimum.

There are no indications for chromospheric activity of GJ\,1214. If similar to the Sun, bright regions should be associated with chromospheric activity, but no - or very weak - line emission has been measured at H$\alpha$, Ca K and in the UV iron lines \citep{Youngblood2017}.

To summarize, there is no independent indication for the presence of bright regions in the photosphere of GJ\,1214. Therefore we consider it as rather unlikely that bright regions could have been dominating over dark spots in modifying the transmission spectrum. In the following, we present an alternative explanation for the transmission spectrum measured by \cite{Rackham2017}.

Assuming a flat optical planetary spectrum (strongly suggested by the bulk of published transmission spectroscopy results), a potential explanation for the measurement of a negative $\Delta k$ is the occultation of spots \citep{Herrero2016}. Spot crossing events during transits, if not properly modeled or removed, will produce negative results for $\Delta k$, especially in the blue optical wavelengths, where the contrast of the spots is higher \citep{Pont2013,Oshagh2014,Herrero2016}. Although our long-term photometry provides a spot filling factor for the times of the transits of \cite{Rackham2017}, the surface distribution of spots are degenerated in latitude. Therefore, we cannot realistically simulate spot occultation events for these specific transits.

Instead, we simulated and analyzed all the transits that occurred during seasons 2013 and 2014, which also included the transit events observed by \cite{Rackham2017}. We did this without removing the spot crossing features, in order to produce the statistics of $\Delta k$ variations when the projected filling factor of spots is 0.05 - 0.1 and the spot crossing features are not properly removed from the transit photometry. The results from our analysis of the 278 transits simulated for seasons 2013 and 2014 are displayed in the bottom panel of Fig.~\ref{GJ1214_tdv}, showing the spurious negative slope toward the blue wavelengths produced by spot crossing features. Overplotted are the transmission spectroscopy measurements of \cite{Rackham2017}, which fall within the region of minimum and maximum $\Delta k$ as derived from the simulations. 

No spot crossing signatures are photometrically resolved in the transit light curves of \cite{Rackham2017}. Our transit simulations indicate that their photometric accuracy is sufficient to rule out the full occultation of a spot of the same size as the planet or larger. However, the complexity of the surface map during these seasons, showing a number of small active regions, could make small spot crossing features undetectable in the observed light curves. As shown in Fig.~\ref{GJ1214_map}, the surface of GJ1214 exhibited a complex activity pattern during the years 2013 and 2014 (BJD from 56300 to 57000), consisting of big persisting spot groups and a large number of small spots spread over a wide range of longitudes. The analysis of the transit depth variations from our simulations show that the deviations in the measurements of $k$ due to spot crossing events can be of up to $5 \cdot 10^{-3}$, when three to four small spots are occulted. In the analyzed transit simulations, such small spots produce a relative flux increase in-transit of below $3 \cdot 10^{-3}$ in B filter, while the transits observed by \cite{Rackham2017} show residuals of up to $5 \cdot 10^{-3}$ in relative flux in the blue wavelengths. Therefore, such transit observations could include small spot crossing features producing a $\Delta k$ of $-3 \cdot 10^{-3}$ as observed by \cite{Rackham2017}. That could explain the observed chromatic trend in transit depth. However, we emphasize again that no spot crossing event was photometrically detected in the \cite{Rackham2017} transits. Therefore, the presented option of small crossed star spots can only be one viable possibility, rather than the final explanation of the measured transmission spectrum.

The case of the three transit observations of \cite{Rackham2017} illustrates the difficulty to measure the slope in a planetary transmission spectrum if the host star is very active. Our long-term photometry, in principle, allows for a correction of unocculted spots. However, we interpret the \cite{Rackham2017} measurements as being affected by unocculted and occulted spots at the same time. In this case, the solution is highly degenerate and no correction of the transmission spectrum for spots is possible anymore. The only solution would be a higher photometric precision per transit measurements to photometrically resolve the in-transit star spot crossing features. Unfortunately, typical ground-based transit observations in low-resolution spectroscopy do not reach the photon-noise limited precision \citep{MallonnStrass,Huitson2017}. Many authors observe multiple transit events of an individual system to verify the robustness of results and increase the precision of the derived planetary transmission spectrum \citep{Lendl2016,Gibson2017}. However, without sufficient S/N per individual data set, this approach does not necessarily increase the accuracy of the transmission spectrum because it might not allow us to differentiate effects of unocculted and occulted star spots by the photometric resolution of spot crossing features. Especially for M dwarfs the photometric detection tend to be burdensome. Their star spots often exhibit a low temperature contrast  and produce only small spot bumps at red wavelengths. At blue wavelengths, where the spot bump is somewhat larger, their photometric resolution is difficult because of the low number of photons emitted by the host star.

In the case of photometrically resolved spot bumps, we suggest the simultaneous fitting of a transit model, the in-transit spot bump as well as correlated noise. Examples of capable software tools are \texttt{spotrod} \citep{Beky2014}, \texttt{PRISM} \citep{TregloanReed2013}, and \texttt{SOAP-T} \citep{Oshaghsoapt}. Recently, \cite{Juvan2017} developed \texttt{PyTranSpot}, which allows for the simultaneous analysis of multicolor light curves of active stars.

The presence of faculae has a very strong impact on the radial velocity signature, both on amplitude and shape, as they are regions where convection is blocked. \cite{Herrero2016} was able to constrain the facula-to-spot area ratio for HD\,189733 to be close to zero by the analysis of simultaneous photometric and radial velocity monitoring. We suggest the observation of similar data for GJ\,1214 to constrain $Q$.

\section{Conclusion}
\label{sec_conc}

We presented a photometric monitoring campaign over five seasons of GJ\,1214 as part of the host stars monitoring survey VAMOS. In the observed period, the mid M dwarf, which hosts a super-Earth planet, showed significant photometric variations due to active regions rotating in and out of view. The data reveal a clear signal in the calculated periodogram at $125\,\pm\,5$~days, which we interpret as the stellar rotation period. Our result does not confirm previous measurements of the rotation period. The period of GJ\,1214 is not unusual for a fully convective M dwarf of 0.176 solar masses. When compared to other northern MEarth targets from \cite{Newton2016a}, it matches with the upper envelope of the mass-period distribution. GJ\,1214 is among the most variable objects of the slow rotators, confirming the result of \cite{Newton2016a} of no correlation between rotation period and photometric variability amplitude for star of lower mass than 0.25 solar masses.

We confirm the finding of \cite{Newton2017} that low-mass stars, such as GJ\,1214, without chromospheric emission have periods longer than 80 days. Previous claims on a shorter rotation period of GJ\,1214 contradicted this result. \cite{Newton2017} also described a positive correlation of photometric variability with chromospheric line emission. New spectroscopy of H$\alpha$ would be needed to verify if the measured increase in photometric variability is accompanied by an increase in chromospheric emission when compared to the non-detection of activity by \cite{Charbonneau2009}.

We modeled the five-year multicolor photometry with the light curve inversion tool \texttt{StarSim} \citep{Herrero2016} and derived the color dependent photometric zero points, defining the relative flux level of an unspotted photosphere. We find that GJ\,1214 was on average dimmed by star spots by 6\,\% in Johnson B and Johnson V, and by 4\,\% in Cousins I. Furthermore we derived a star spot contrast temperature of 370~K, in agreement with previous results of weaker spot contrasts for late type stars than for solar type stars \citep{Berdyugina2005}. The variation is stable in phase over four years from 2013 to 2016, being caused by long-living active longitudes. The values of the longitudes remain stable over time within the measurement precision, thus we find no indications for differential rotation in the stellar photosphere. We measured an increase in amplitude of photometric variability over the time interval covered by our observations. Also the light curve inversion with \texttt{StarSim} yields an increasing spot filling factor. This can be interpreted as an activity cycle similar to the 11-year cycle of the Sun. The existence of such cyclic behavior in fully convective stars is still an open question. The data presented in this work point toward a cycle length of more than four
years. If we compare the photometry of this work with MEarth photometry of the years 2008 to 2010, the variability roughly monotonically increased from 2008 to 2016, being indicative of a cycle length of more than eight years.

From the \texttt{StarSim} modeling we were able to derive a surface map of the stellar photosphere for all 635 transit epochs between the beginning and end of our five seasons of monitoring. We used these maps to produce the associated transit light curve from 400 to 2000~nm with \texttt{StarSim}. These light curves, all affected by a certain amount of spots, were fitted with a transit light curve model. This exercise resulted in values for the mean, minimum and maximum star spot modification of the transit depth by unocculted star spots over wavelength. In principle, this exercise provides the wavelength-dependent correction of transit depth for unocculted spots for any transit epoch during the observing campaign. 

The effect of unocculted spots on the transit depth is always positive, meaning that the depth of transit is enhanced. \cite{Rackham2017} observed three transit events in 2013 and 2014 and measured a slightly shallower transit depth at blue optical wavelengths than previous studies. Our photometry shows their transit epochs to be close to photometric minima, making it rather unlikely that the shallow transit depth is caused by unocculted bright regions. Another viable explanation for a shallower transit depth are occulted dark spots hidden in the photometric noise of the transit light curve. However, despite having simultaneous long-term photometry on hand, we cannot correct for occulted spots when their signature is not photometrically resolved. This makes it impossible to break the degeneracy in the optical slope of a derived transmission spectrum between spot effects and planetary spectral signature for active host stars.

\begin{acknowledgements}
E.H., A.R. and I.R. acknowledge support by the Spanish Ministry of Economy and Competitiveness (MINECO) and the Fondo Europeo de Desarrollo Regional (FEDER) through grant ESP2016-80435-C2-1-R, as well as the support of the Generalitat de Catalunya/CERCA program. I.G.J. acknowledges the Austrian Forschungsf\"orderungsgesellschaft FFG project ``RASEN'' P847963. This research has made use of the SIMBAD data base and VizieR catalog access tool, operated at CDS, Strasbourg, France, and of the NASA Astrophysics Data System (ADS).
\end{acknowledgements}

%
\bibliographystyle{aa} 
\bibliography{GJ1214_bib} 
%

\end{document}